# Durable, ultrathin, and antifouling polymer brush coating for efficient condensation heat transfer

Shuai Li[1], Cheuk Wing Edmond Lam[2#], Matteo Donati[2], Kartik Regulagadda[2], Emre Yavuz[1], Till Pfeiffer[3], Panagiotis Sarkiris[4], Evangelos Gogolides[4], Athanasios Milionis[2], Dimos Poulikakos[2*], Hans-Jürgen Butt[1], Michael Kappl[1*]

1. Max Planck Institute for Polymer Research, Mainz, Germany.

2. Laboratory of Thermodynamics in Emerging Technologies, Department of Mechanical and Process Engineering, ETH Zurich, Zurich, Switzerland.

3. Institute for Technical Thermodynamics, Technical University of Darmstadt, Germany.

4. Institute of Nanoscience and Nanotechnology, NCSR “Demokritos”, Attiki, Greece

E-mail: dpoulikakos@ethz.ch; kappl@mpip-mainz.mpg.de

**Abstract**

Heat exchangers are made of metals because of their high heat conductivity and mechanical stability. Metal surfaces are inherently hydrophilic, leading to inefficient filmwise condensation. It is still a challenge to coat these metal surfaces with a durable, robust and thin hydrophobic layer, which is required for efficient dropwise condensation. Here, we report the non-structured and ultrathin (~6 nm) polydimethylsiloxane (PDMS) brushes on copper that sustain high-performing dropwise condensation in high supersaturation. Due to the flexible hydrophobic siloxane polymer chains, the coating has low resistance to drop sliding and excellent chemical stability. The PDMS brushes can sustain dropwise condensation for up to ~8 h during exposure to 111 °C saturated steam flowing at 3 $m \cdot s^{-1}$, with a 5-7 times higher heat transfer coefficient compared to filmwise condensation. The surface is self-cleaning and can reduce bacterial attachment by 99%. This low-cost, facile, fluorine-free, and scalable method is suitable for a great variety of condensation heat transfer applications.

## 1. Introduction

Water vapor condensation is ubiquitous in nature and everyday life. [1, 2] It plays an important role in a variety of applications involving heat and mass transfer, [3-5] e.g., for water harvesting, [6-8] water desalination, power generation, and thermal management. Most heat transfer devices are manufactured from metals with high thermal conductivity, e.g. ~398 $W \cdot m^{-1} \cdot K^{-1}$ for copper. However, the metals are hydrophilic and easily wetted by condensate from steam, leading to a stable liquid film covering the surface. [9, 10] During this so-called "filmwise condensation" mode, the liquid film hinders heat transfer because of its significant thermal resistance. By applying a low-adhesion or hydrophobic polymer coating on the metal surface, [11-13] the condensate can nucleate, grow, coalesce, and easily slide away from the surface in the form of distinct droplets. This condensation mode is called "dropwise condensation". [14] It can show a performance enhancement of up to one order of magnitude compared to filmwise condensation thanks to the periodic condensate removal, which leaves an accessible surface for fresh droplet nucleation. [4, 15]

On the other hand, these polymeric coatings usually have very low thermal conductivities on the order of 0.1 - 0.5 $W \cdot m^{-1} \cdot K^{-1}$. [16] A thick polymer coating increases thermal resistance, leading to an inefficient heat transfer process. For example, the state-of-the-art coatings, e.g., superhydrophobic surfaces, [17-20] and lubricant-infused surfaces, [21-25] usually have a large thickness ranging from micrometer to millimeter (Figure 1a, TableS1-S2, Supporting Information). It can lead to a significantly high thermal resistance of more than $10^{-5}$ $m^2 \cdot K \cdot W^{-1}$ on copper substrate, [26-28] compromising the heat transfer benefits from the dropwise condensation mode. To reduce the thermal resistance, ultrathin polymer brushes (ideally at the nanoscale level), such as polydimethylsiloxane (PDMS) brushes,[29-36] can be grafted onto the metal substrate. The coatings are ultrathin (~6 nm) with low thermal resistance (< $10^{-7}$ $m^2 \cdot K \cdot W^{-1}$, Figure S1, Supporting Information) and able to repel water drops with low contact angle hysteresis (< 10°).

Achieving a small coating thickness usually comes at the cost of compromised robustness. Despite the nanoscale thickness, PDMS brushes are promising alternative materials compared to superhydrophobic and lubricant-infused surfaces due to the absence of micro- or nanoscale rough surface topography, which typically is prone to damage,[37] as well as the good adhesion to the substrate due to strong covalent grafting. On the contrary, for superhydrophobic surfaces, the superhydrophobicity relies on vapor cushions within the micro/nanostructures (Cassie state).[38-40] At elevated supersaturation, impalement of the micro/nanostructures by water will occur (Wenzel state),[41] thus the surface loses its superhydrophobicity, leading to filmwise condensation (Figure 1b).[42, 43] For lubricant-infused surfaces, although studies have shown their excellent liquid repellency, and heat transfer coefficient up to 5 times higher compared to filmwise condensation,[26] they still face the issue of gradual lubricant depletion in the long term (Figure 1b).[44]

Another problem in condensation applications is the contamination on the surface, e.g., biofouling, which can be a major issue that limits the heat transfer efficiency in industrial applications, e.g., condenser tubes,[45, 46] and heat exchangers.[47] Microorganisms, such as bacteria, can attach to the surface of the condenser fins and tubes, acting as defects,[48] and continue expanding to form a fouling layer that can affect the heat transfer. This fouling layer can reduce the efficiency of the condenser by acting as an insulator, increasing the resistance to heat transfer and obstructing the departure of water. Although this problem is more relevant to the cooling side (internal part of condenser tubes), it is not rare that the external part of the condenser tubes faces the problem of contamination. For example, in atmospheric water harvesting applications, [49, 50] dust or microorganisms may attach to the surface during environmental exposure. Under ambient conditions, bacteria can easily grow and form a fouling layer. Therefore, antifouling is a very desirable property of coatings for heat transfer applications. Finally, the green chemistry of the polymeric coating material is also essential. With continuous use, coating degradation is inevitable, and bio-persistent elements can be released into the environment, especially in processes that involve open systems. Specifically, hydrophobic surfaces sustaining dropwise condensation are usually made with long-chain perfluorinated polymers, which are not environmentally friendly and their byproducts during degradation tend to bioaccumulate. [51]

Here, we study the condensation of water on PDMS brushes (Figure 1c) under harsh experimental conditions. Because of their strong covalent bond with the substrate (Figure S2, Supporting Information) and the absence of rough surface microfeatures, PDMS brushes are stable even at challenging, high subcooling values and steam pressures. We experimentally demonstrate the coating resilience with an accelerated endurance test characterized by exposure to superheated steam at 111 °C and 1.42 bar with a shear velocity of 3 $m \cdot s^{-1}$. Under the aforementioned conditions, the PDMS brushes can sustain dropwise condensation for at least 8 h and show a 5 times greater heat transfer coefficient compared to filmwise condensation. We also show that the PDMS brushes can effectively repel bacteria such as *Escherichia coli* and *Staphylococcus aureus*, reducing the attachment by 99%. With all these merits, PDMS brushes are promising to open a new avenue to enhance practical heat transfer performance by sustainable and effective means.

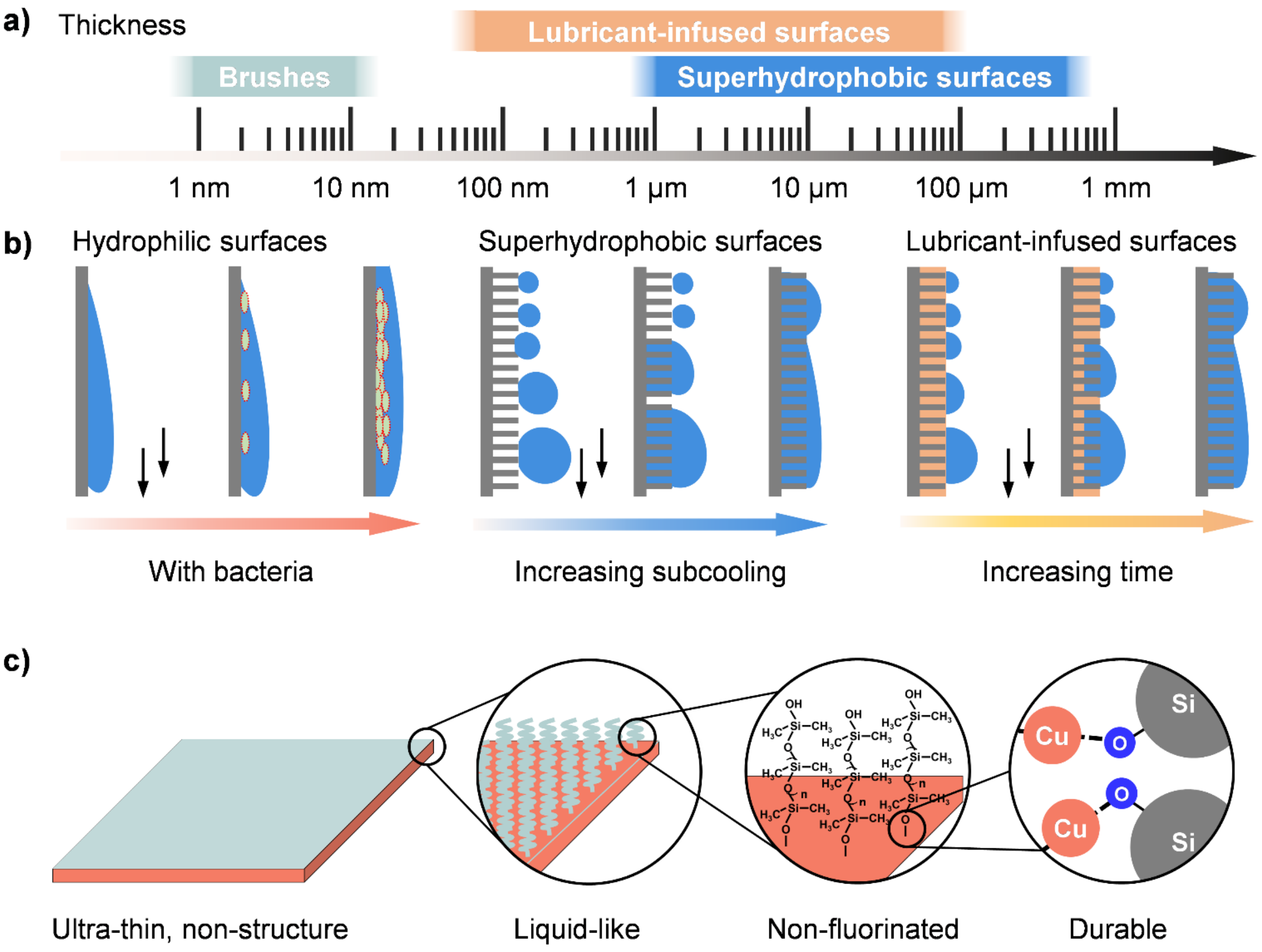

**Figure 1.** Overview. a) Summarized coating thickness of brushes, superhydrophobic surfaces, and lubricant-infused surfaces. b) Schematic showing the limitations of hydrophilic surfaces, superhydrophobic surfaces, and lubricant-infused surfaces with the accumulation of bacteria, increasing subcooling, and increasing time, respectively. c) Surface characteristics of PDMS brushes.

## 2. Results and discussion

### 2.1 Preparation and characterization of PDMS brushes.

PDMS brushes were prepared by a drop-casting and annealing method as described by Krumpfer & McCarthy for silicon wafers (Figure 2a and Methods). [52-56] Briefly, a PDMS liquid drop (molecular weight of 11740 $g \cdot mol^{-1}$) was deposited onto an oxygen-plasma-activated copper surface, followed by heating. The grafting process of PDMS brushes onto the surface is initiated by siloxane hydrolysis. Then the silanol-terminated chain can be covalently bonded onto the hydroxyl group on the copper (Figure S3, Supporting Information). [29, 57, 58] The resulting PDMS coating on copper is smooth with root-mean-square roughness of 3.6 nm ± 0.5 nm in an area of 500 × 500 $nm^2$ (Figure 2c). This value is mainly related to the roughness of the pristine copper substrate (3.0 nm ± 0.7 nm). The cost of the coating is estimated to be less than 10 USD per $m^2$ (Table S3, Supporting Information).

The coated surface is hydrophobic with water advancing contact angle of 111° ± 1° and a contact angle hysteresis of 10° ± 4° (Figure 2b). PDMS brushes could also be easily applied on a variety of materials, such as silicon and aluminum, leading to similarly improved wetting properties: water advancing contact angle and contact angle hysteresis on silicon and aluminum substrates are 108° ± 1°, 8° ± 1°, and 110° ± 1°, 12° ± 1°, respectively. In addition, PDMS brushes can be applied on curved surfaces. As shown in Figure 2d and Video S1 (Supporting Information), a water drop slides off on a PDMS-coated cylinder copper surface (diameter: 24 mm) within 1 s.

The thickness and grafting density of PDMS brushes were analyzed by atomic force microscopy (AFM) force spectroscopy (Figure S4, Supporting Information), giving a coating thickness of $d$ = 6 nm ± 1 nm. It is reported recently that lubricant thickness could be optimized for efficient condensation process on lubricant-infused surfaces utilizing drop coarsening due to merging by lateral capillary forces. [59] Such a mechanism cannot work for the much thinner PDMS brush coatings, but their orders of magnitude smaller thickness leads to a negligible thermal resistance, which is even more beneficial for heat transfer. Following the equation $\Gamma = (d\rho N_A)/M_w$, the grafting density $\Gamma$ of our PDMS brushes was be estimated as 0.3 ± 0.05 chains·nm$^{-2}$, where $\rho$ and $N_A$ represents mass density and Avogadro constant, respectively. [60]

The water condensation dynamics on PDMS brushes are first studied at the microscale, *in situ*, using an environmental scanning electron microscope (ESEM, FEI Quanta 650 FEG). As shown in Figure 2e, the droplets maintain spherical cap shapes. While growing, the droplets easily coalesce without visible contact line pinning, suggesting excellent droplet mobility and water repellency of PDMS brushes for tiny condensing droplets. When considering superhydrophobic surfaces, it can be challenging to repel tiny droplets because the superhydrophobicity relies on the empty space within structures. [19, 61, 62] At high subcooling values, if the droplet size is comparable to or smaller than the size of the surface features, droplets may stay pinned inside these features, and this may cause coalescence at early growth stages with other droplets and consequently lead to the formation of filmwise condensation. [39-43, 63]

As a primary durability test for the PDMS brushes, we used drop sliding measurement in a custom-built device, as reported before. [64] A needle connected with a peristaltic pump generated series of water drops (each 45 μL). The stage was tilted at 50° and a high-speed camera (FASTCAM MINI UX100, see Methods for details) was attached to the stage to capture videos. After continuously sliding thousands of water drops over the surface, PDMS brushes still exhibit good hydrophobicity (Video S2-S5, Supporting Information). The velocity of drop 1 and drop 5000 was 0.08 m·s$^{-1}$ ± 0.02 m·s$^{-1}$ and 0.10 m·s$^{-1}$ ± 0.02 m·s$^{-1}$, respectively (Figure S5, Supporting Information). The corresponding dynamic advancing contact angle and contact angle hysteresis for drop 1 and drop 5000 were 130° ± 3°, 66° ± 7°, and 126° ± 3°, 56° ± 6°, respectively. The larger value of dynamic contact angle hysteresis compared to the static contact angle hysteresis shown in Figure 2b is attributed to the substantially higher drop velocity. [65]

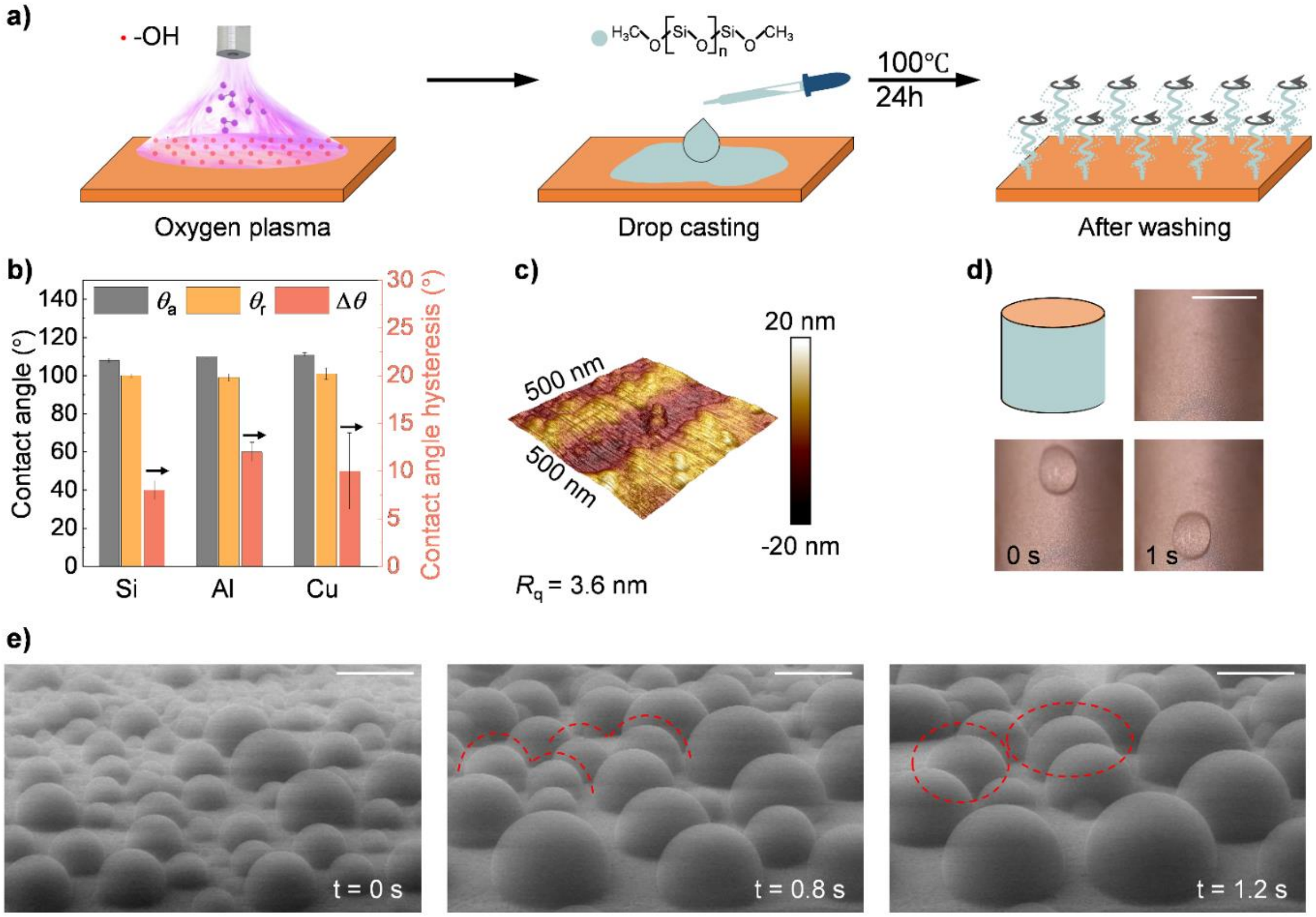

**Figure 2.** Preparation and characterization of PDMS brushes. a) Fabrication of PDMS brushes via drop casting method. b) Wetting properties (advancing and receding contact angles, contact angle hysteresis) of water on PDMS brushes-coated silicon wafer, aluminum, and copper substrates. c) Surface morphology of PDMS brushes on copper substrate. d) Time-lapsed photographs of a water drop sliding on a PDMS brushes-coated copper cylinder (diameter: 24 mm). Tilt angle: 25°. Scale bar: 10 mm. e Time-lapsed photographs of water condensation and drop coalescence on PDMS brushes by environmental scanning electron microscopy. Scale bar: 20 μm.

## 2.2. Condensation heat transfer performance at low pressure.

The condensation heat transfer performance of PDMS brushes was tested with a custom-built condensation chamber under low saturation vapor pressure (30 mbar, steam temperature 24 °C) (Figure S6, Supporting Information, and methods for details). [5, 26] These conditions are comparable to industrial condensers' operation parameters. The steam was generated from an electric boiler and flows horizontally across the sample, where the flow speed was ~4.6 $m \cdot s^{-1}$.

With increased subcooling, the dropwise condensation mode on PDMS brushes was maintained (Figure 3a, Video S6-S8, Supporting Information), without a change of the circular drop shape. On our superhydrophobic reference surface (see Experimental Section for details), [66] dropwise condensation was observed at subcooling <1 K. Superhydrophobic surfaces are known for their jumping dropwise condensation mode only at low subcooling. [17, 67] However, the drops show an irregular shape at subcooling of 2 K and finally turn into a liquid film at 3 K. The filmwise condensation is due to the flooding at high subcooling values, because the surface remains superhydrophobic after condensation upon drying. Filmwise condensation on the superhydrophobic surface still allows higher heat transfer compared to the conventional hydrophilic surface. The reason is the difference in wetting situation. On the conventional hydrophilic surface, a thick liquid film is formed, leading to large thermal resistance. It can be recognized from the bulge formed at the bottom of the hydrophilic surface (Figure 3a, Video S6). On the superhydrophobic surface, filmwise condensation leads to a flooding of the surface structure, which then acts as a wicking layer. Therefore, film thickness is reduced to the order of the structure thickness of the superhydrophobic surface layer and no bulge is visible at the lower end. This leads to a lower thermal resistance of the superhydrophobic layer during filmwise condensation (Figure 3a, Video S7). The different condensation mode for these three surfaces highlight the ability of PDMS brushes to sustain dropwise condensation over a wide range of subcooling values.

This stability of dropwise condensation is also reflected in the trend of heat transfer coefficient as a function of subcooling (Figure 3b). Although the superhydrophobic surface shows better heat transfer performance than PDMS brushes at low subcooling, the performance on the superhydrophobic surface decreases and approaches that for filmwise condensation around a subcooling of 2-3 K. Figure 3c plots the heat fluxes of the three surfaces. At subcooling of 3K, PDMS brushes exhibit a heat flux of 233 $kW \cdot m^{-2} \cdot K^{-1}$, which is 20% higher than that of the superhydrophobic surface. It should be noted that we cannot measure the heat transfer coefficient at higher subcooling (>3K) for these better-performing surfaces due to their efficiency. Nevertheless, it is reasonable to predict that PDMS brushes can still maintain dropwise condensation at higher subcooling values or condensation rates, due to the absence of micro and nanostructures that can eventually get flooded with water.

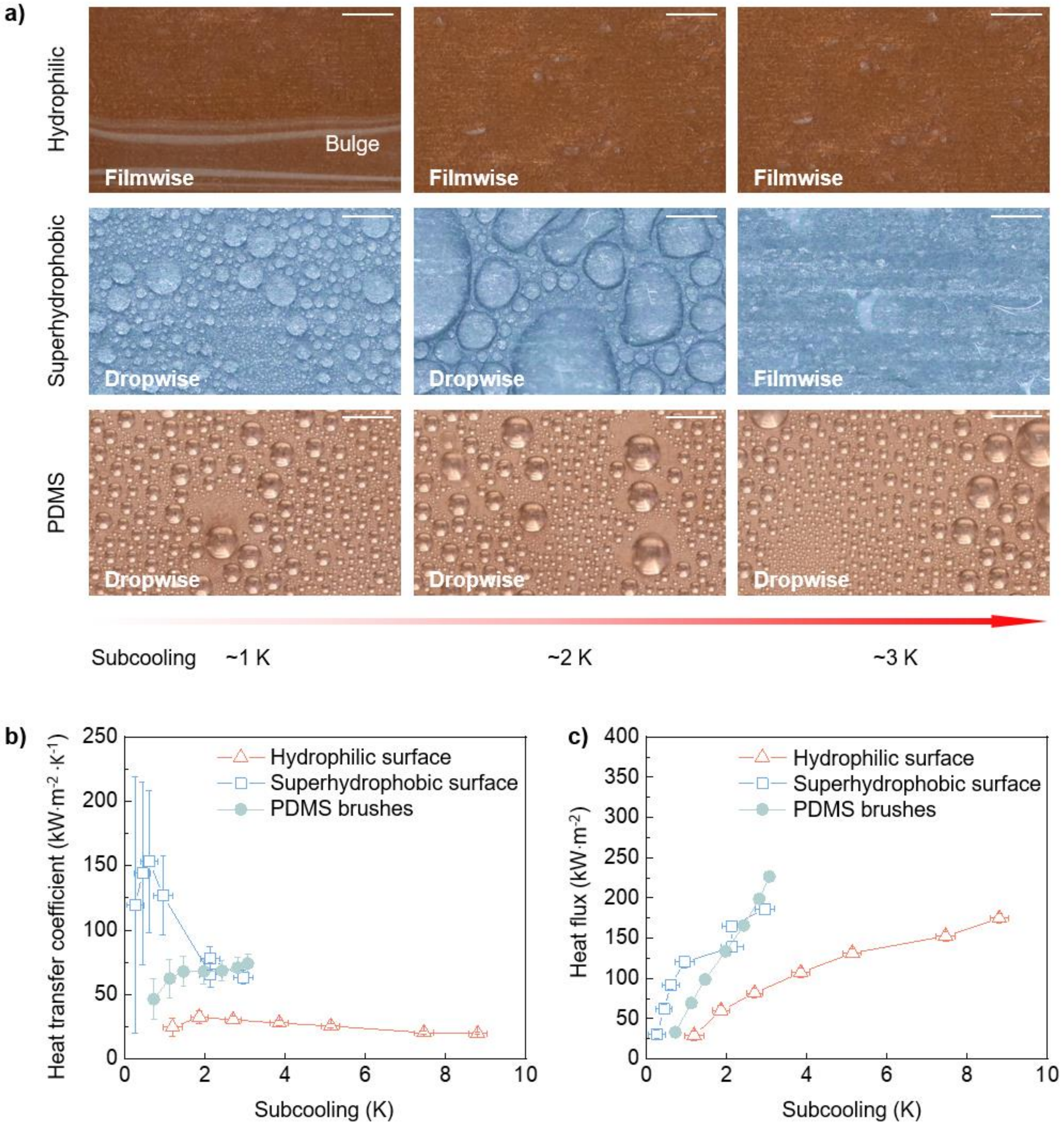

**Figure 3.** Condensation heat transfer performance at low steam pressure (30 mbar). a) Optical images of hydrophilic surface, superhydrophobic surface, and PDMS brushes surface at different subcooling (1K, 2K, and 3K) under steady-state condensation. Scale bar: 2 mm. b) Heat transfer

coefficients of the three vertically placed surfaces at different subcooling. Steam pressure: 30 mbar; steam flow rate: ~4.6 $m \cdot s^{-1}$; steam flow direction: horizontal. c) Corresponding heat flux of the three surfaces at different subcooling. Data for superhydrophobic surface from our recent work. [66]

### 2.3. Condensation heat transfer performance at high pressure.

To quantitatively evaluate condensation heat transfer performance in harsh conditions, the PDMS brush surface was tested in a high-pressure flow chamber where the steam pressure and temperature were 1.42 bar and 111 °C, respectively (Figure S7, Supporting Information). [5] The experiment is conducted in a flow condensation environment while the steam flowed vertically with velocities of 3 $m \cdot s^{-1}$ or 9 $m \cdot s^{-1}$. As shown in Figure 4a,b, PDMS brushes exhibited a higher heat transfer coefficient at both steam velocities. At 3 $m \cdot s^{-1}$, the average heat transfer coefficient reached 125 $kW \cdot m^{-2} \cdot K^{-1}$, which is ~5 times higher than that on a bare CuO reference surface (filmwise condensation). Due to enhanced advection, the heat transfer performance was better on both surfaces at the steam flow rate of 9 $m \cdot s^{-1}$. On PDMS brushes, the heat transfer coefficient reaches 233 $kW \cdot m^{-2} \cdot K^{-1}$, which is ~7 times higher than that on the CuO surface.

To test the coating durability under condensation, we focus on the PDMS brushes with a steam flow rate of 3 $m \cdot s^{-1}$ for an extended period (~8 h). The heat transfer coefficient and corresponding subcooling were continuously measured over 8 h while condensation rates were recorded at several intervals (Figure 4c and Video S9, Supporting Information). In the first 0.5 h the system had to stabilize. The heat transfer coefficient increased and oscillated initially. After ≈0.5 h, the experimental conditions were stable, and the surface exhibited a heat transfer coefficient of ~ 121 $kW \cdot m^{-2} \cdot K^{-1}$. Up until 7 h, dropwise condensation remained the dominant mode. Afterward, an increase of the departure droplet sizes was observed and localized filmwise condensation islands appeared (~30% area shows filmwise condensation). However, the heat transfer coefficient remained as high as 103 $kW \cdot m^{-2} \cdot K^{-1}$ at 8.8 h, which is still more than 4 times higher than that of filmwise condensation. Such accelerated durability test proves that the ultrathin PDMS brushes sustains efficient dropwise condensation in harsh conditions for ~8 h, showing its potential for practical applications where the conditions are much milder. [68] The appeared filmwise condensation in the end may be related to the oxidation process of copper, which degrades the wettability of PDMS coating. [67, 69] It should be noted that even after degradation, the wetting properties of the coating can be restored by applying a bit of PDMS oil (Figure S8, Supporting Information). Moreover, as a perspective of future impacts, although PDMS brush coating is grafted on the flat substrate here, it may also be used to modify structured surface to further enhance condensation. [70]

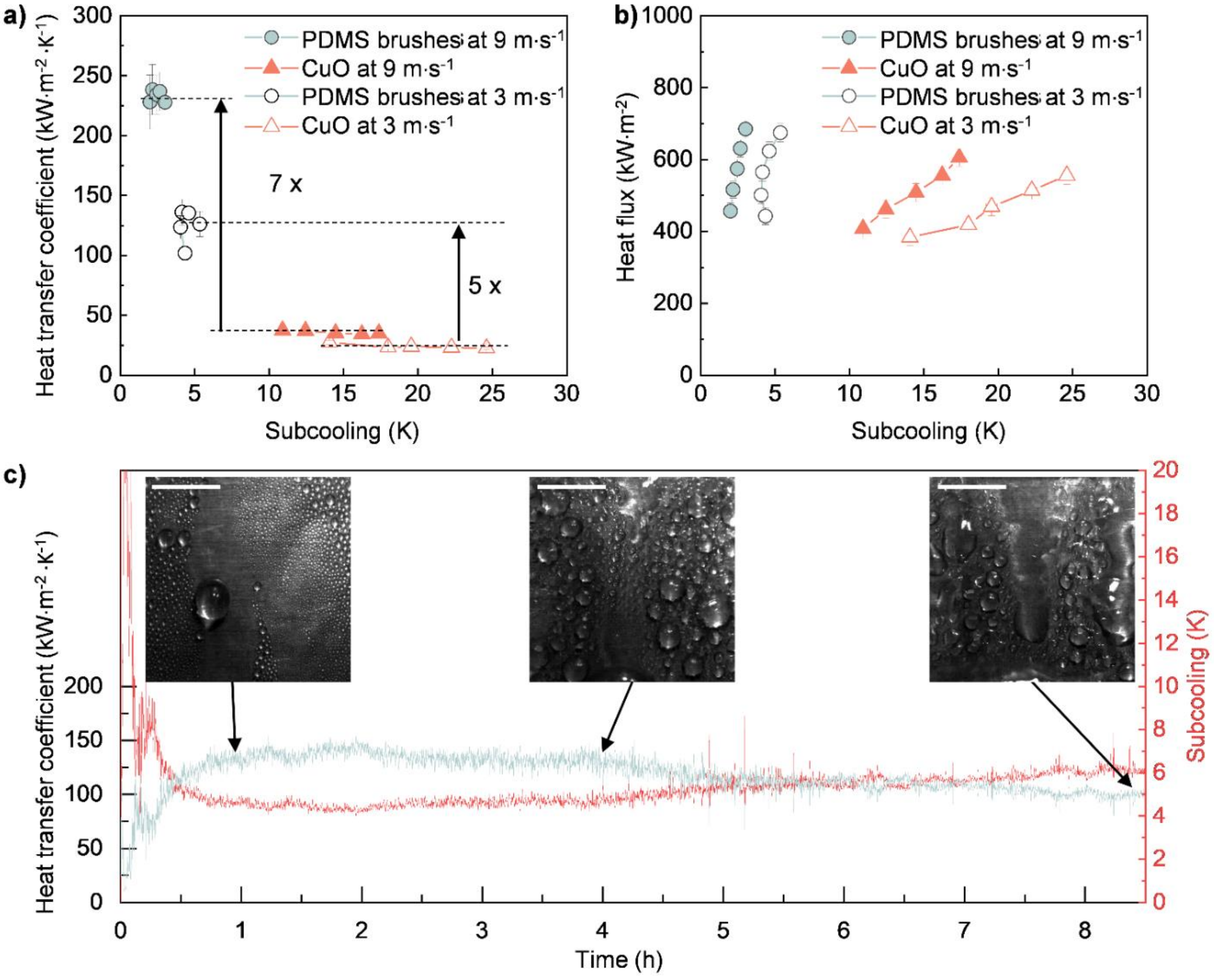

**Figure 4.** Condensation heat transfer performance at high pressure. a) Heat transfer coefficients of vertically placed PDMS brushes and hydrophilic CuO at different subcooling values. Steam temperature: 111 °C; steam pressure: 1.42 bar; steam flow rate: 3 m·s$^{-1}$ and 9 m·s$^{-1}$. Dashed lines mark average values. b) Corresponding heat flux of the two surfaces at different subcooling. c) Heat transfer coefficient of PDMS brushes within ~8 h and corresponding subcooling values. Inset: High-speed photographs of condensed drops. Scale bar: 5 mm.

## 2.4 Antifouling test.

We further measured the antifouling property by immersing the substrate into solutions containing *Escherichia coli* (*E. coli*) and *Staphylococcus aureus* (*S. aureus*), respectively. Both *E. coli* and *S. aureus* are commonly found bacteria. *E. coli* is Gram-negative and rod-shaped, while *S. aureus* is Gram-positive and spherically shaped. After 1 day of culture at 37 °C, the samples are taken out and washed gently to remove the floating bacteria. Scanning electron microscope (SEM) images showed a significantly reduced bacterial number on PFDTS and PDMS surfaces when compared to those on the uncoated surface (Figure 5a, Figure S9, Supporting Information). Specifically, the attached *E. coli* number density was $(2.7 \pm 0.3) \times 10^{11}$ $m^{-2}$ on Si surfaces, $(1.5 \pm 1.1) \times 10^{9}$ $m^{-2}$ on PFDTS surfaces, and $(2.6 \pm 1.0) \times 10^{9}$ $m^{-2}$ on PDMS surfaces (Figure 5b). For *S. aureus*, the number density on the three surfaces were $(9.1 \pm 4.9) \times 10^{11}$ $m^{-2}$, $(1.1 \pm 0.8) \times 10^{10}$ $m^{-2}$, and $(1.0 \pm 0.6) \times 10^{10}$ $m^{-2}$, respectively. The calculated anti-bacterial efficiency (i.e., the ratio of reduced bacterial amount on the surface to the total amount on Si surface) reached ~99% on both PFDTS and PDMS surfaces, showing the comparable antifouling property of PDMS surface to the conventional fluorinated PFDTS surface. A quick anti-contamination test showed that the PDMS brushes can effectively repel adhesive materials such as chalk powder and chili sauce, revealing its self-cleaning property (Figure S10 and Video S10-S11, Supporting Information).

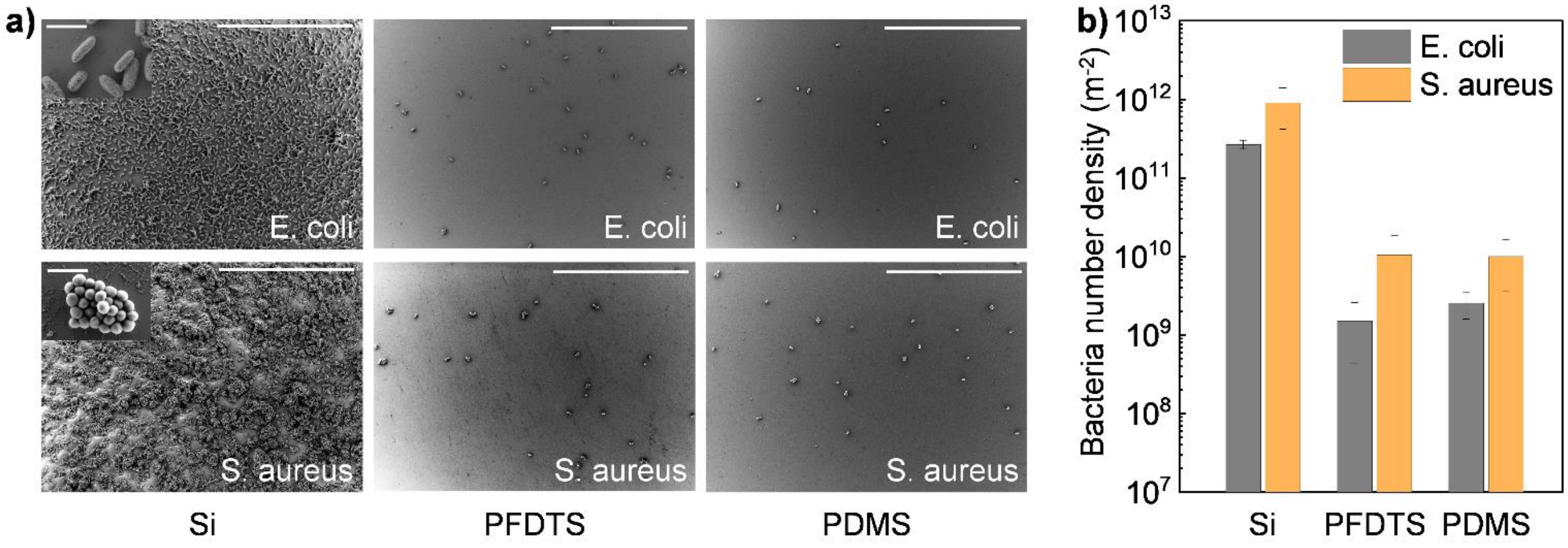

**Figure 5.** Antifouling performance of PDMS brushes. a) SEM images of *E. coli* (rod-shaped, top three images) and *S. aureus* (spherical-shaped, bottom three images) on silicon wafer, PFDTS, and PDMS brushes surface. Scale bar: 50 μm, and 2 μm (inset). b) Number density of attached bacteria on three surfaces.

## 3. Conclusion

We demonstrate that the low cost, flat, antifouling, and non-fluorinated PDMS brushes coating can sustain high-performing dropwise condensation at extreme conditions, e.g., high subcooling value, and high steam shear flow and temperature. The PDMS brushes consists of siloxane polymer chains, where one side is covalently grafted onto the substrate and the other side is free and flexible. The coating is thin (thickness of 6 nm) and water-repellent (advancing contact angle ~110°, contact angle hysteresis of ~10° on copper). The experimental results show that PDMS brushes on copper substrate exhibit dropwise condensation and ~3-7 times higher heat transfer coefficients compared to that of filmwise condensation formation on pristine copper substrates in the low (30 mbar) and high (1.4 bar) pressure chambers. The PDMS brushes also exhibit excellent durability in high-pressure environment, which is confirmed by the 8-hour condensation test under harsh conditions of 1.4 bar steam pressure and 3 $m \cdot s^{-1}$ steam velocity. Additionally, PDMS brushes can effectively repel 99% bacteria, e.g., *E. coli* and *S. aureus*. Therefore, PDMS brushes are promising candidates for developing low-cost, environment-friendly, and effective coating for condensation heat transfer applications.

## 4. Experimental Section

*Surface preparation*: PDMS brushes are prepared by drop-casting. First, the substrates (silicon, aluminum, or copper) were washed in acetone, isopropanol, and deionized water with ultrasonication for 10 min, respectively. Then they are treated with an oxygen plasma (Diener Electronic Femto, 120 W, 6 $cm^3 \cdot min^{-1}$ oxygen flow rate) for 5 min. Afterwards, several drops of PDMS (polydimethylsiloxane, 100 cSt, Thermo Scientific) are applied on the substrate, which is then covered by spontaneous wetting. After full spreading the substrates were put in the oven at 100 °C for 24 hours and washed with acetone afterward to remove any unbound residue. This preparation method is repeated twice. PFDTS surfaces are prepared via chemical vapor deposition in a vacuum desiccator. 20 µL 1H,1H,2H,2H-perfluorodecyltrimethoxysilane is added before the desiccator is vacuumed below 20 mbar. The reaction lasts for 4 hours. Hydrophilic CuO surfaces are prepared by immersing oxide-free pristine copper in boiling water for 30 minutes. The superhydrophobic surfaces are fabricated as described before.[66] Briefly, after the cleaning process, the substrates are immersed into a 9.25% V/V aqueous solution of hydrochloric acid for 10 minutes to fabricate microstructures. Then they are immersed into boiling water for 5 minutes to fabricate nanostructures on top of microstructures through the boehmitage process. Finally, the substrates are coated with a thin hydrophobic film through $C_4F_8$ plasma deposition.

*Surface characterization:* Advancing and receding contact angles are measured using a goniometer (OCA35, Dataphysics). The volume of a sessile water was gradually (1 µL·s$^{-1}$) increased from 5 µL to 20 µL and the decreased back to 5 µL. The contact angles were determined by fitting an ellipse to the contour images. Surface morphology was measured using Dimension Icon AFM (Bruker) in tapping mode. Reflective Aluminum Si cantilevers (OLTESPA-R3) with a spring constant of ~2 N·m$^{-1}$ were used. The thickness of the brush layer was measured using an AFM (JPK Nanowizard 4) in force mapping mode. Force−distance curves were recorded with a grid of 16 × 16 points on an area of 1 × 1 µm$^2$. For observation of condensation using environmental scanning electron microscope (ESEM) (Quanta 650 FEG, FEI), the samples were placed on a custom-made copper platform, which was cooled with a recirculating chiller and maintained at ~2 °C. The drop velocity and dynamic contact angles on the surface were determined by analyzing videos of drop sliding via a MATLAB program (DSAfM). The videos were recorded using a high-speed camera (FASTCAM MINI UX100, Photron with a Titan TL telecentric lens, 0.268×, 1″ C-Mount, Edmund Optics) at a frame rate of 500 FPS. Briefly, the edge position of the drops were detected, after the drop images were corrected by subtracting the background from the original images and tilted according to the background image. The drop velocities were calculated by the displacement from each frame. Dynamic advancing and receding contact angles were determined by applying a 4th-order polynomial fit to the drop contour in each image.

*Condensation heat transfer measurements:* We used two custom-made experimental setups for the condensation tests similar to our recent work. [5, 26] The chambers are evacuated firstly before the introduction of steam. An electric boiler was used to generate steam from deionized water. Condensation tests were performed with saturated steam at a pressure of 1.42 bar or 30 mbar. The samples were mounted on the copper block. Several temperature sensors inside the copper block were used to determine the condensation heat flux ($q''$) through the surface by following the equation $q'' = k_c A_c / A_e \cdot dT/dx$. Here, $k_c$ is the thermal conductivity of the copper cooler, $A_c$ is the cross-sectional area of the cooler, $A_e$ is the area of the exposed condensing surface, and $dT/dx$ is the constant thermal gradient along the array. $dT/dx$ was computed from a least-squares linear fit of the temperatures measured with the temperature sensor array. In the low-pressure (30 mbar) chamber, the surface temperature was measured by two temperature sensors attached to the surface. Videos were recorded with a DSLR (D7500, Nikon) and a macro lens (AF Micro-Nikkor 200mm f/4D IF-ED, Nikon). In the high-pressure chamber (1.42 bar), the surface temperature was estimated using a thermocouple placed inside the substrate. The videos were recorded using a high-speed camera (FASTCAM Mini UX100, 2000FPS) and the same lens. More details are in Figure S6-7, Supporting Information.

*Antifouling tests:* To test the antifouling property, the samples (1x1 $cm^2$) are firstly sterilized by UV light (366 nm) for 15 min. Then the samples are placed in a sterile 24-well plate, and each well included 2 mL of the bacteria test suspension (refer Figure S11 in Supporting Information for the preparation of bacterial suspension). The samples are incubated for 1 day at 37 °C, before the medium is removed from the samples and gently washed 3 times with 1 mL saline solution (0.85% NaCl in Milli-Q water). Afterward, the bacteria is fixed by 1 mL glutaraldehyde (Sigma-Aldrich, 2.5% (v/v) in the saline solution) for 30 min at room temperature. Subsequently, the coatings are gently washed 3 times with the saline solution and dehydrated with a series of ethanol (30, 40, 50, 60, 70, 80, 90, 95, and 99.89%, 15 min each, last step twice). Lastly, the samples are dried under vacuum at room temperature overnight prior to SEM imaging. For bacterial number counting, more than 30 images are taken at random positions by scanning electron microscopy (SEM, LEO 1530 Gemini, Zeiss).

## Data availability

The authors declare that the data supporting the findings of this study are available within the paper and its Supporting Information or from the corresponding author upon reasonable request.

## Acknowledgments

This project has received funding from the European Research Council (ERC) under the European Union's Horizon 2020 research and innovation programme (HARMoNIC, grant number 801229; Advanced grant DyanMo, No 883631). Shuai Li thanks the China Scholarship Council (CSC) for the financial support. Till Pfeiffer acknowledges the financial support by the German Research Foundation - Project number 265191195 – SFB 1194; subproject C03.

## Author contributions

S. L., A. M., D. P., H.-J. B., and M. K. designed the research and experiments. S. L. and A. M. prepared the surface. S. L. carried out the experiments and characterization unless otherwise stated below. C. W. E. L. conducted the ESEM measurements. C. W. E. L., M. D., and K. R. conducted the condensation heat transfer measurements. P. S. and E. G. prepared the superhydrophobic surface for reference. S. L. and E. Y. conducted the antifouling measurements. S. L., C. W. E. L., M. D., K. R., E. Y., P. S., E. G., A. M., D. P., M. K., and H.-J. B. wrote the manuscript. All authors have given approval to the final version of the manuscript.

## Competing interests

The authors declare no competing interests.

Present Addresses:

[#]C.W.E.L.: Massachusetts Institute of Technology, Cambridge, MA, 02139, United States

# Supporting Information for

# Durable, ultrathin, and antifouling polymer brush coating for efficient condensation heat transfer

*Shuai Li[1], Cheuk Wing Edmond Lam[2#], Matteo Donati[2], Kartik Regulagadda[2], Emre Yavuz[1], Till Pfeiffer[3], Panagiotis Sarkiris[4], Evangelos Gogolides[4], Athanasios Milionis[2], Dimos Poulikakos[2*], Hans-Jürgen Butt[1], Michael Kappl[1*]*

1. Max Planck Institute for Polymer Research, Mainz, Germany.

2. Laboratory of Thermodynamics in Emerging Technologies, Department of Mechanical and Process Engineering, ETH Zurich, Zurich, Switzerland.

3. Institute for Technical Thermodynamics, Technical University of Darmstadt, Germany.

4. Institute of Nanoscience and Nanotechnology, NCSR "Demokritos", Attiki, Greece

E-mail: dpoulikakos@ethz.ch; kappl@mpip-mainz.mpg.de

**This file includes:**

Figures S1 to S11

Tables S1 to S3

Videos S1 to S11

Supporting References

**Supplementary Figures**

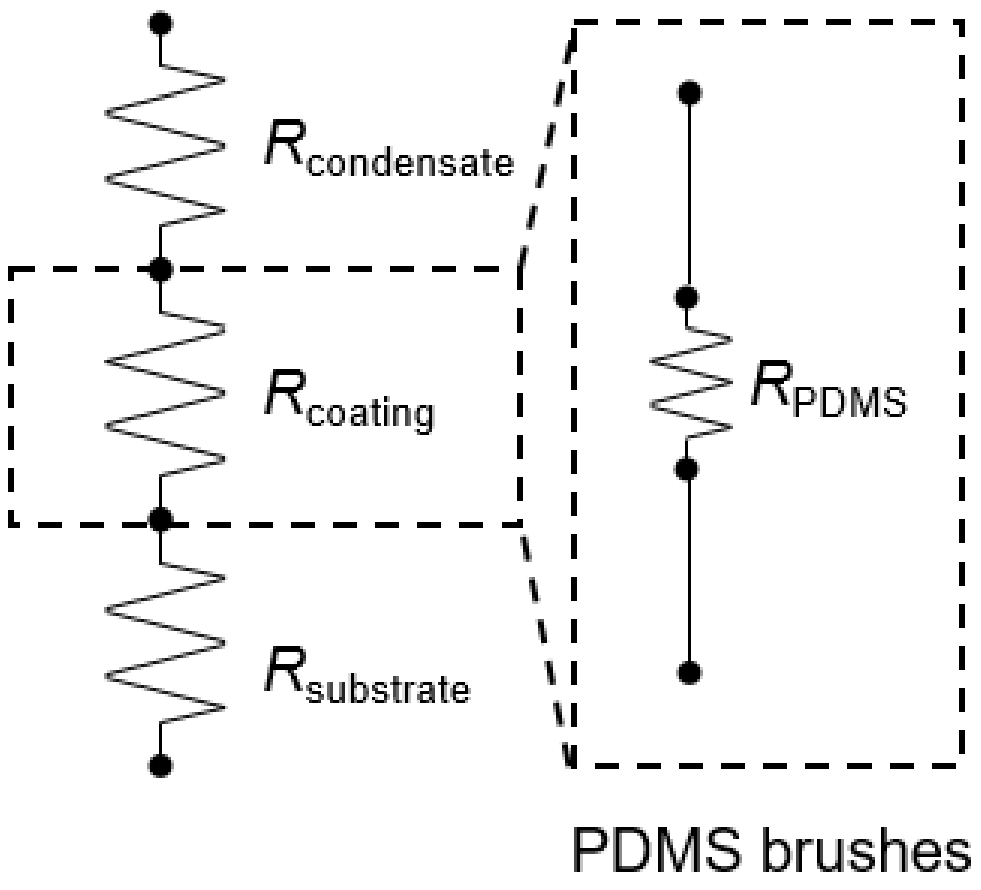

**Figure S1**. Thermal resistance. Schematic of the thermal resistance network of PDMS brushes.

Based on the assumption of one-dimensional heat conduction, thermal resistance ($R$) of three surfaces are calculated.[1-3] For polydimethylsiloxane (PDMS) brushes, thermal resistance is calculated simply using $R_{PDMS} = d_{PDMS}/k_{PDMS}$, where $d_{PDMS}$ and $k_{PDMS}$ is coating thickness and thermal conductivity, respectively. Thermal conductivity PDMS is set to 0.16 W·m$^{-1}$·K$^{-1}$.

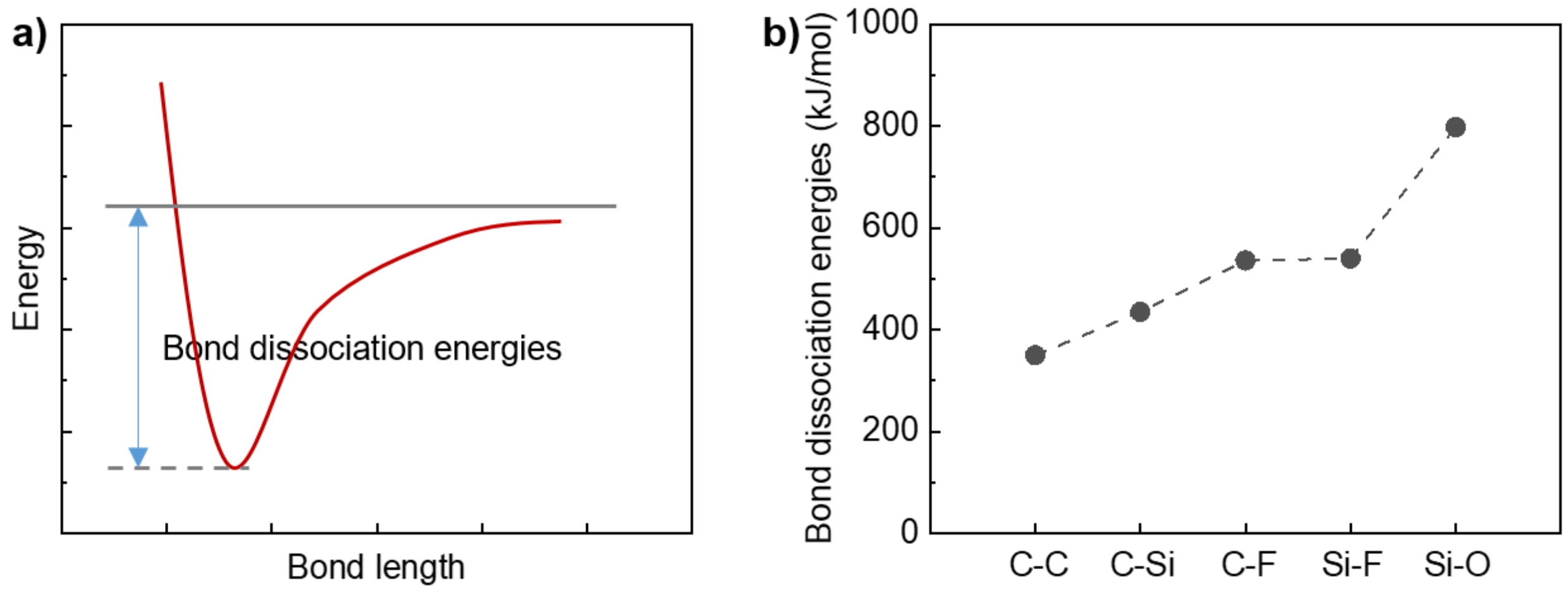

**Figure S2.** Bond dissociation energies. a) Schematic showing the definition of bond dissociation energies: the energy required to break a bond and form two atomic or molecular fragments.[4] b) Comparison of bond dissociation energies for different bonds.[5]

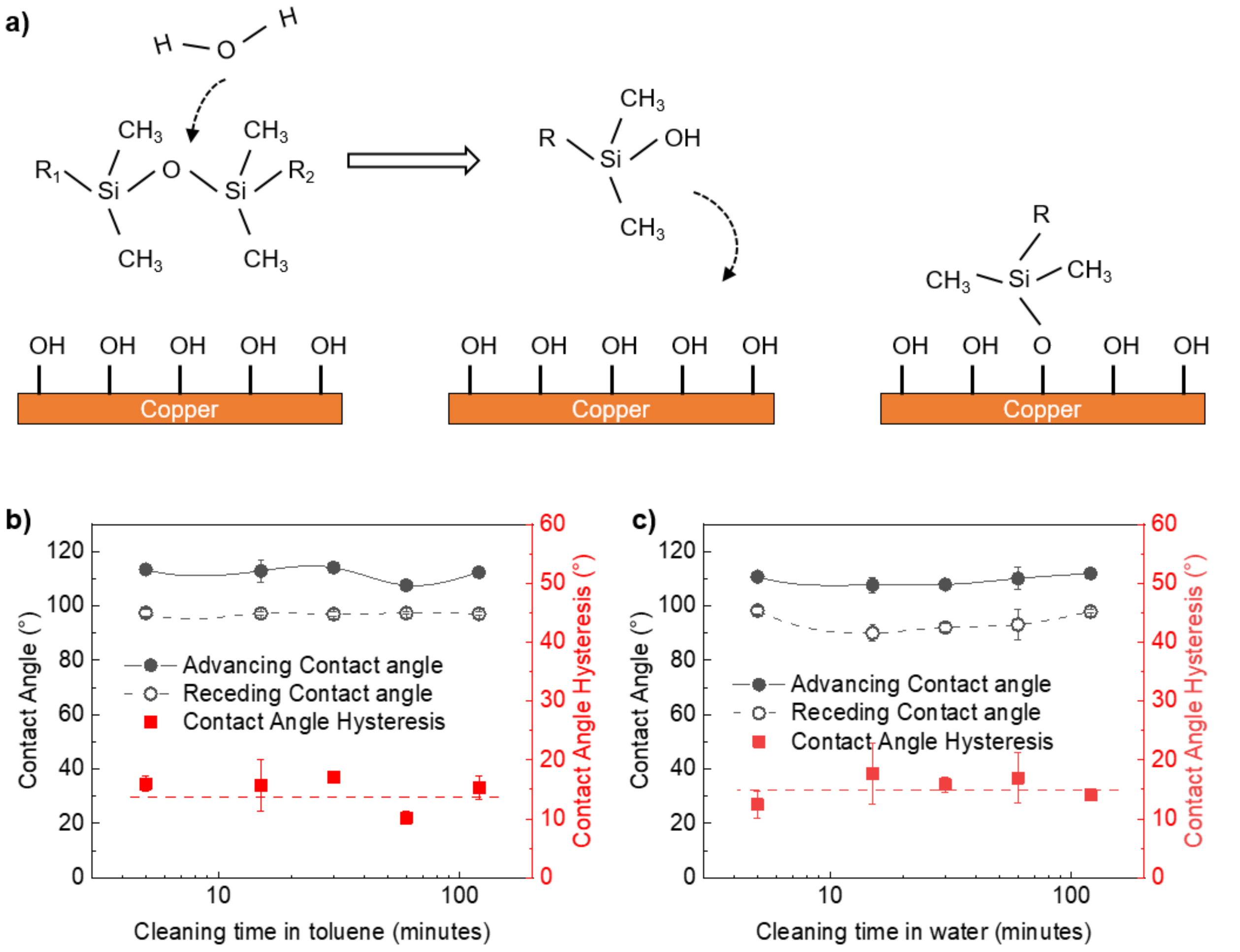

**Figure S3.** a) Schematic showing the bonding process of PDMS brushes on the copper. b) Water advancing contact angle, receding contact angle, and contact angle hysteresis on PDMS brushes after ultrasonic cleaning in toluene for different times. c) Water advancing contact angle, receding contact angle, and contact angle hysteresis on PDMS brushes after ultrasonic cleaning in water for varying time.

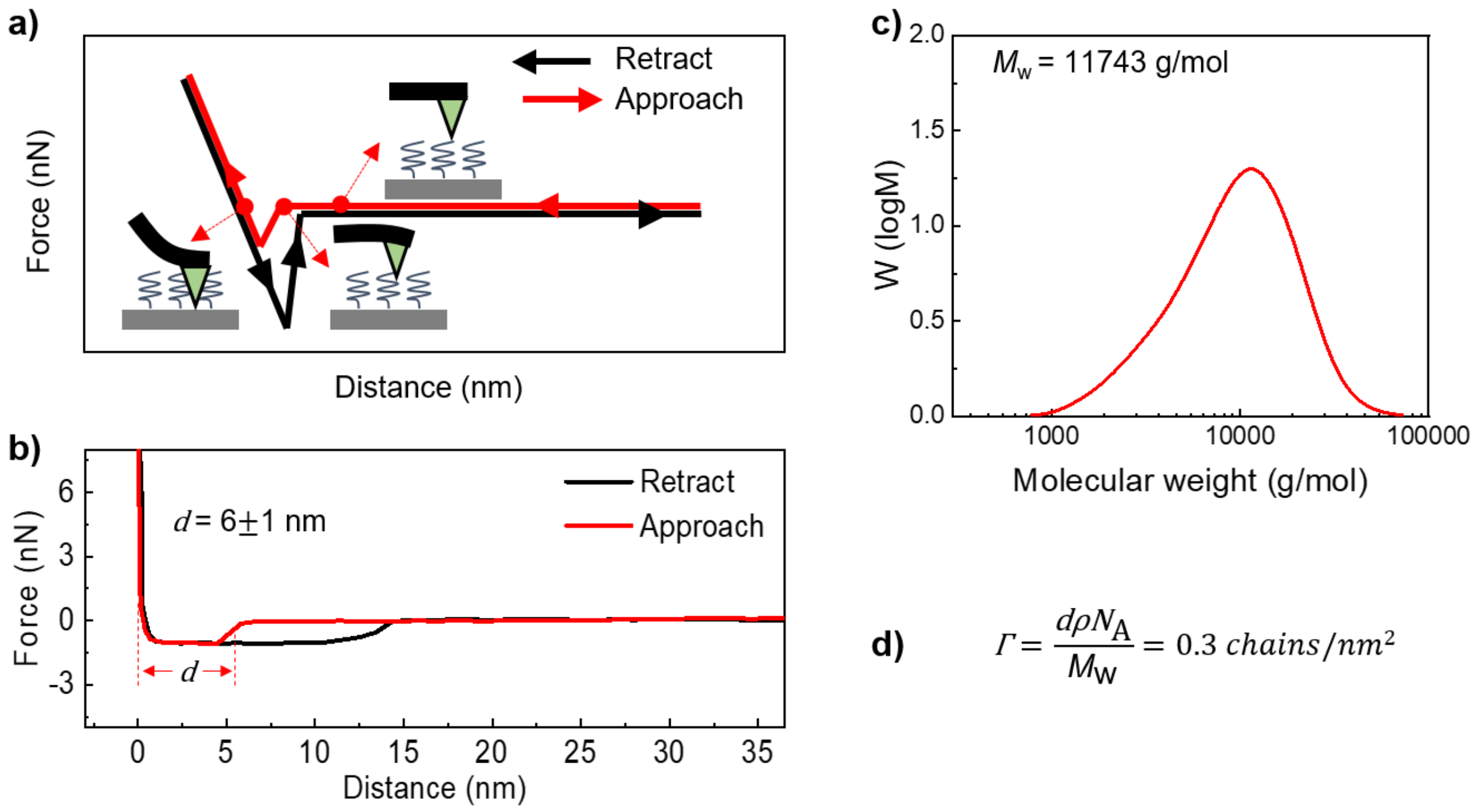

**Figure S4.** Estimation of grafting density of PDMS brushes. a) Schematic showing the state of the AFM cantilever and tip corresponding to different positions during the approach and retract phases during a force curve. b) Representative force curves measured by AFM. The brush thickness is extracted as the distance between onset of attractive force (AFM tip touching the brush surface) and the hard wall repulsion (tip penetrating the brush). The measured brush thickness $d$ is 6 nm ± 1 nm. c) Molecular weight distribution and average molecular weight $M_w$ measured by gel permeation chromatography. d) Equation used for calculation of grafting density $\Gamma$, where $\rho$ and $N_A$ represent mass density and Avogadro constant, respectively.

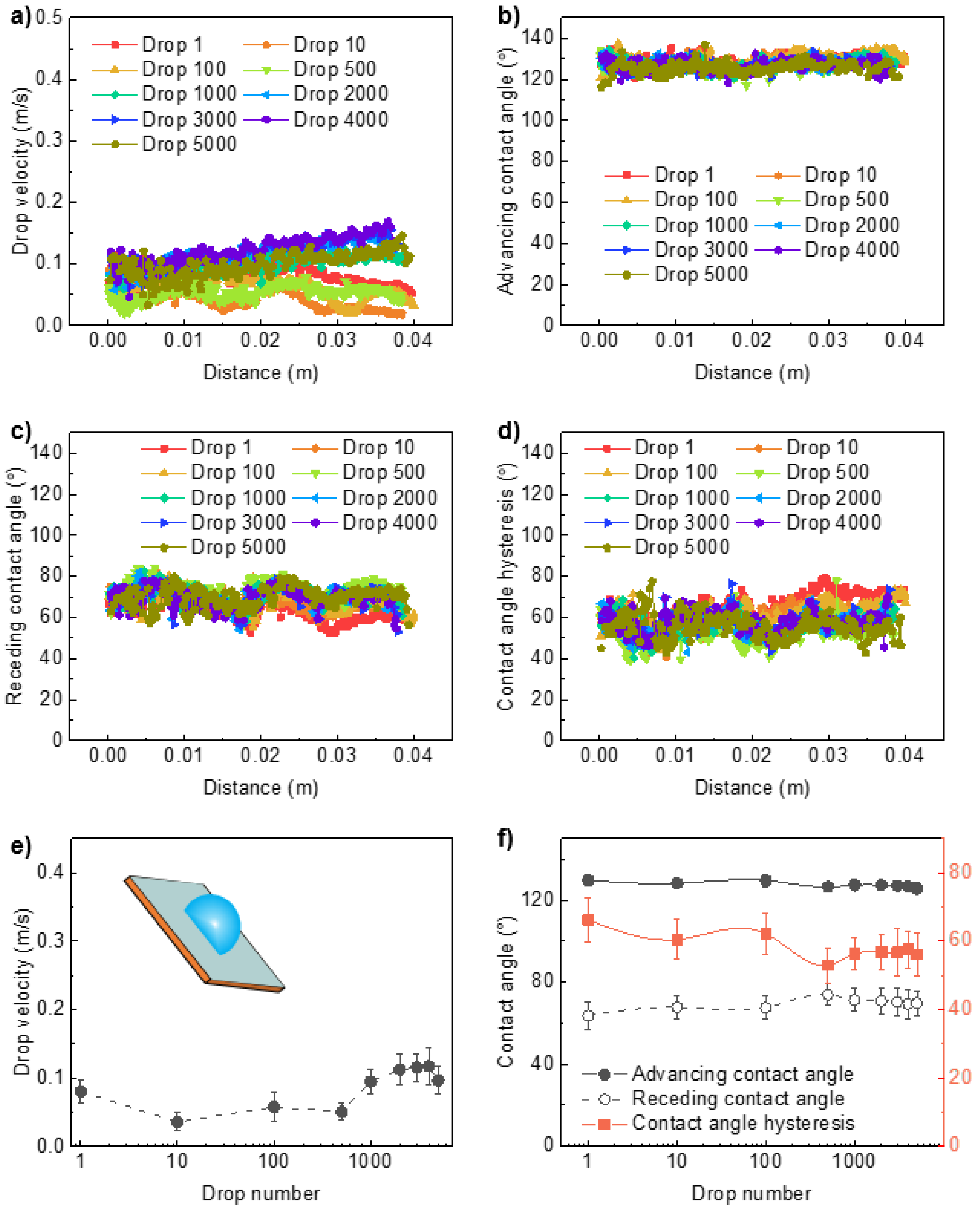

**Figure S5.** Continuous water drop sliding on PDMS brushes coated copper plate. a) Velocity of successive water drops as a function of displacement. b) Advancing contact angle of successive water drops as a function of displacement. c) Receding contact angle of successive water drops as a function of displacement. d) Contact angle hysteresis of successive water drops as a function of displacement. Surface tilt angle: 50°. Drop volume: 45 μL. e) Sliding velocity of successive droplets on tilted PDMS brushes. Surface tilt angle: 50°. Drop volume: 45 μL. Inset: Schematic of water drop slide on the surface. f) Water advancing and receding contact angles, and contact angle hysteresis of successive droplets on tilted PDMS brushes.

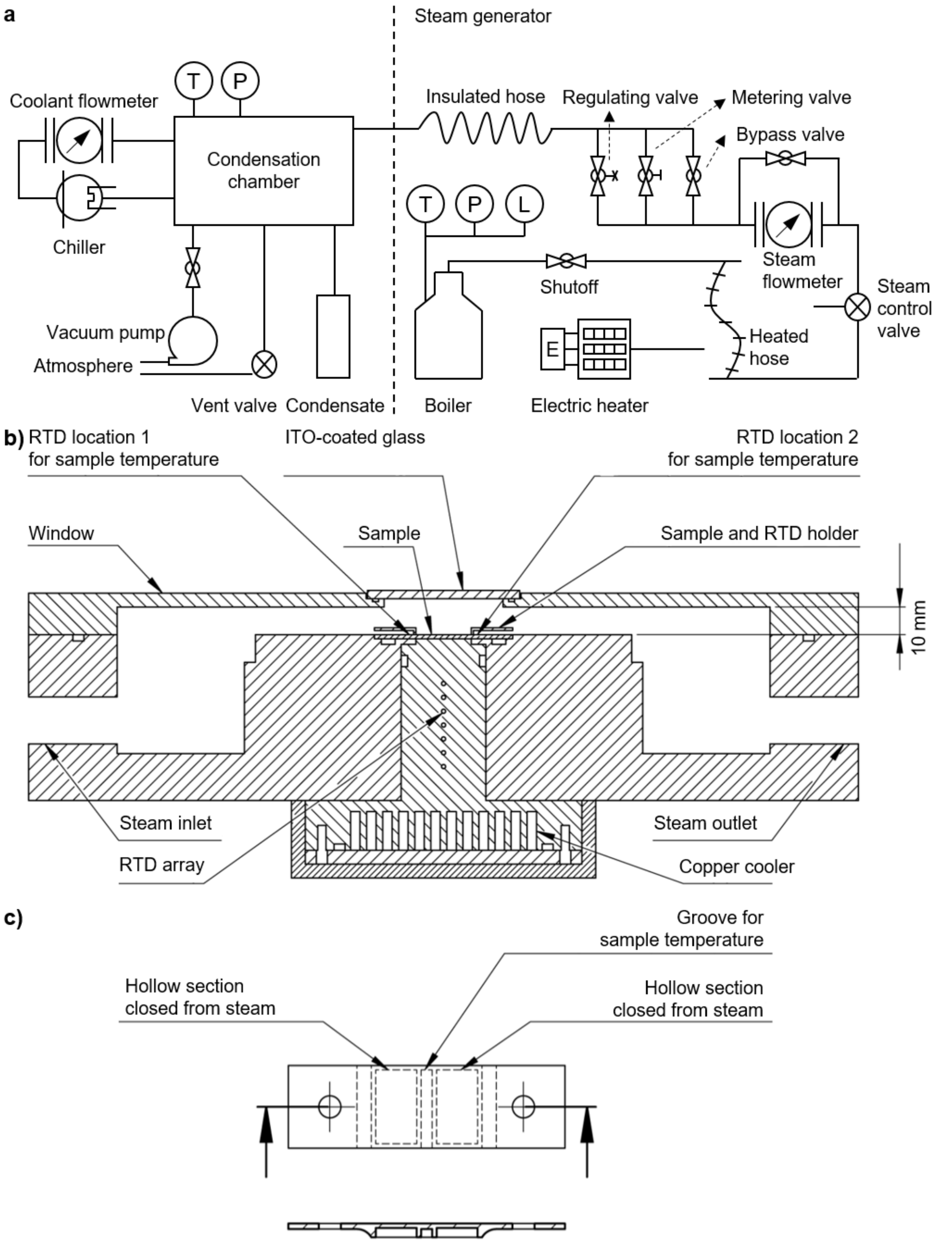

**Figure S6.** Device for condensation heat transfer measurements at 30 mbar. a) Schematic of the overall experimental setup. b) Top-view cross-section of the condensation chamber. c) Schematic of sample and surface temperature RTD holder, top view and cross-section.

For low-pressure heat transfer measurement, we follow similar procedures to our previous work, which is referred to for details.[1] A short summary is provided below. In this current work, low-pressure heat transfer measurement is conducted at 30 mbar.

To minimize non-condensable gases in the testing chamber, deionized water is continuously boiled at > 1.4 bar, draining the steam for 30 minutes and the condensation chamber is vacuumed to < 0.01 mbar (1 Pa) at the same time. During experiments, steam is generated at 1.4 ± 0.01 bar and its flow rate is measured by a flowmeter (FAM3255, ABB). Regulating and metering valves reduce the pressure and temperature of the steam as it enters the chamber to 30 ± 0.5 mbar and ~24 ˚C (corresponding saturation temperature). The steam is at saturation. A vacuum pump (RC 6, VACUUBRAND) maintains a continuous flow of the low-pressure saturated steam horizontally across the surface. The resulting mean steam speed in the channel is ~ 4.6 $m \cdot s^{-1}$. The copper cooler at the back of the test surface is cooled by a recirculating chiller (WKL 2200, LAUDA), where the flow rate is monitored by a flowmeter (SITRANS FM MAG5000 and SITRANS FM MAG 1100, SIEMENS). An array of 7 RTDs is in the cooler to obtain the heat flux through it with a linear fit. In the chamber, the surface temperature is monitored continuously with two RTDs. The chamber steam temperature is similarly monitored with two RTDs. A capacitance gauge is used to monitor the chamber steam pressure. To avoid condensation before chamber steam conditions are stabilized at the said pressure and temperatures, the chiller is set to an initial coolant temperature of 25 °C, higher than the target steam temperature. Once steam is stabilized, the coolant temperature is reduced to trigger condensation at 7 set points. As the system reaches steady state at each chiller set point, the coolant flow rate is set to 180 ± 10 $L \cdot h^{-1}$ and measurements are taken over 1 minute, during which the chamber pressure has to maintain at 30 ± 0.5 mbar and the boiler pressure at 1.4 ± 0.01 bar without intervention. Then the coolant temperature is lowered to cover the remaining set points to obtain heat fluxes and heat transfer coefficients at different subcoolings.

Heat transfer coefficient is defined as described in the main text. Its computation, and the uncertainty propagation procedure can be seen in our previous work.[1]

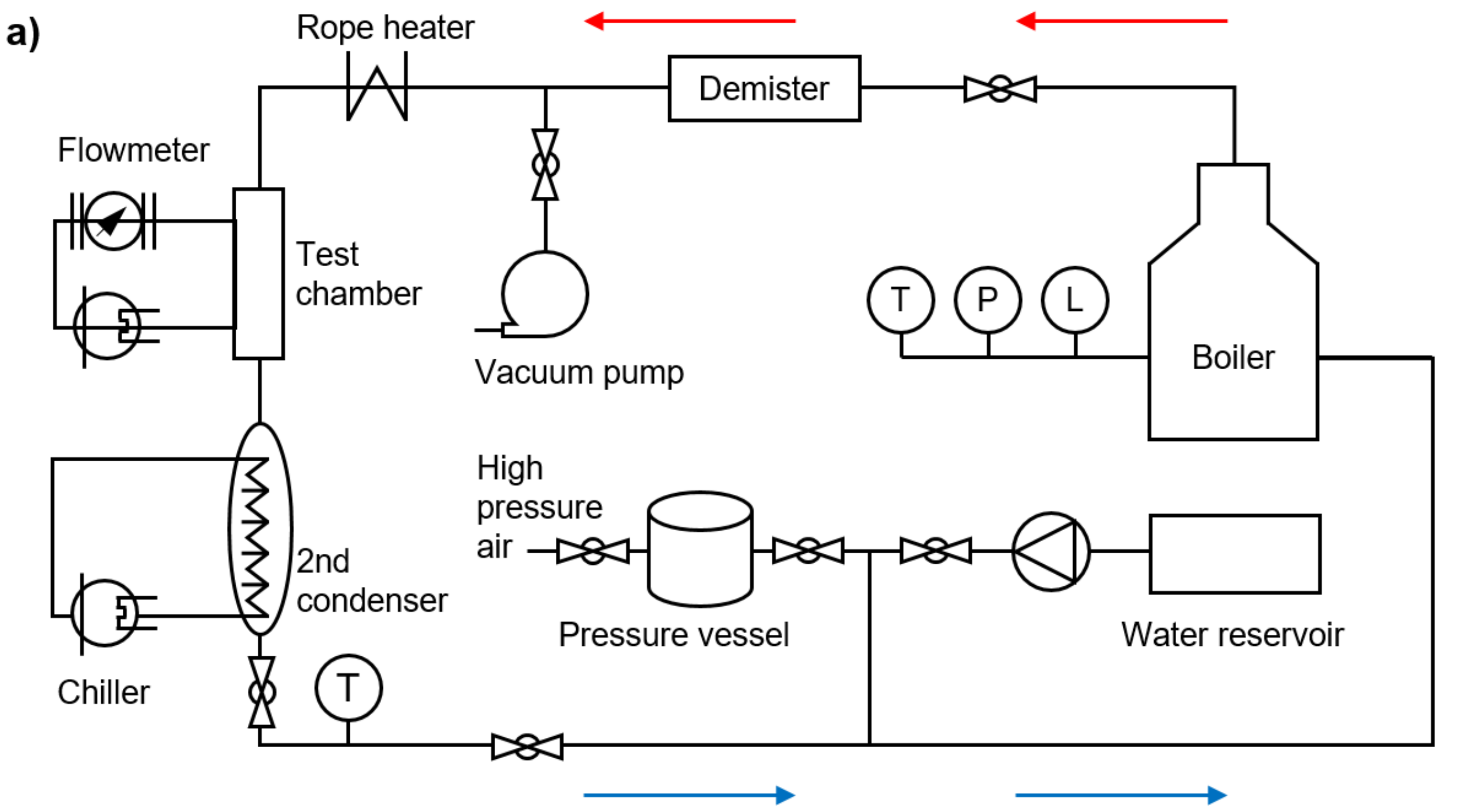

**Figure S7.** Device for condensation heat transfer measurements at 1.4 bar. a) Schematic of the overall experimental setup. For high-pressure heat transfer measurement, we follow similar procedures to our previous work,[6] which is referred to for details. The high-pressure flow chamber is installed into a loop so that experiments can be run continuously for an extended period. To fully degas the water and exclude the non-condensable gases, the condensation chamber is initially vacuumed to below 20 mbar. After that, the whole setup it is filled with liquid water by opening the connection with a water reservoir. To further reduce the amount of non-condensable gases, water is repeatedly pumped and a valve at the highest point of the setup is repeatedly opened to release the air residues. During this process the pressure in the setup is always kept above the atmospheric pressure. For surface temperature, a thermocouple is installed into the substrate. During experiments, the steam pressure is kept at 1.42 bar and the mean steam speed in the channel is controlled at 3 $m \cdot s^{-1}$ or 9 $m \cdot s^{-1}$.

b)

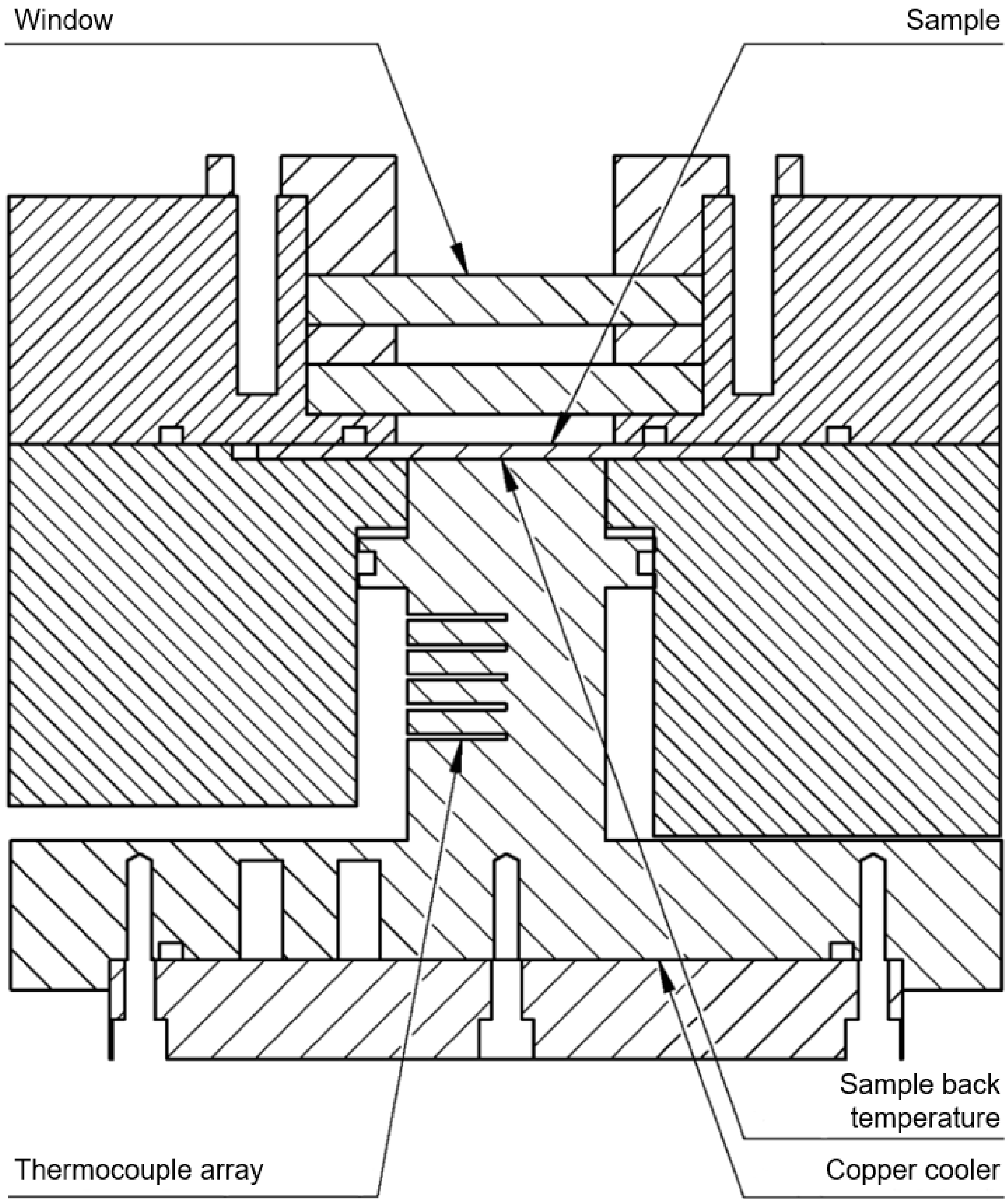

**Figure S7.** Device for condensation heat transfer measurements at 1.4 bar. b) Schematic of the condensation chamber, top-view cross-section. Heat transfer coefficient is defined as described in the main text. Its computation, and the uncertainty propagation procedure can be seen in our previous work.[6]

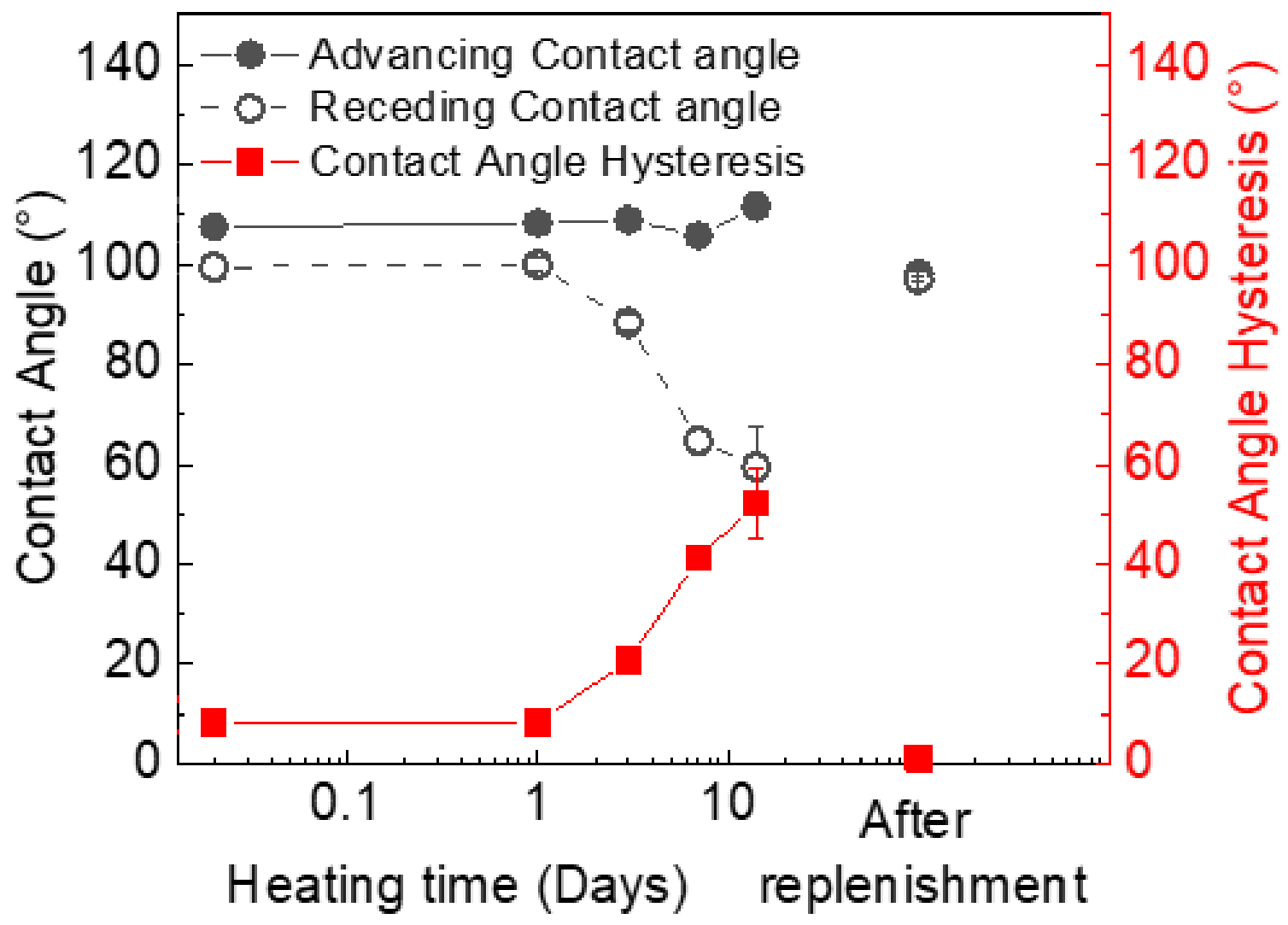

**Figure S8.** Durability test of PDMS-coated copper in hot water (~100 °C) and its recovery. After several days in hot water, the surface lost its wetting property towards water with a high contact angle hysteresis. After replenishment by applying a bit of PDMS oil on top, low contact angle hysteresis is achieved again.

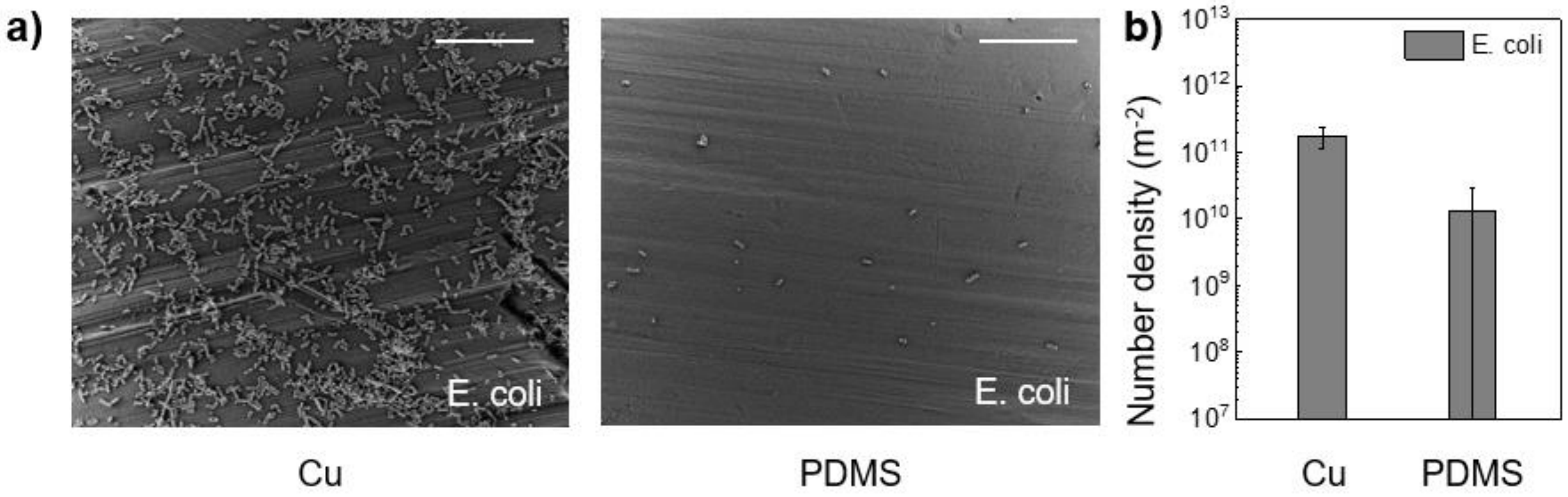

**Figure S9.** Antifouling property of different surfaces. a) SEM images of E. coli on pristine copper and PDMS-coated copper. Scale bar: 20 μm. b) Number density of attached bacteria on the two surfaces.

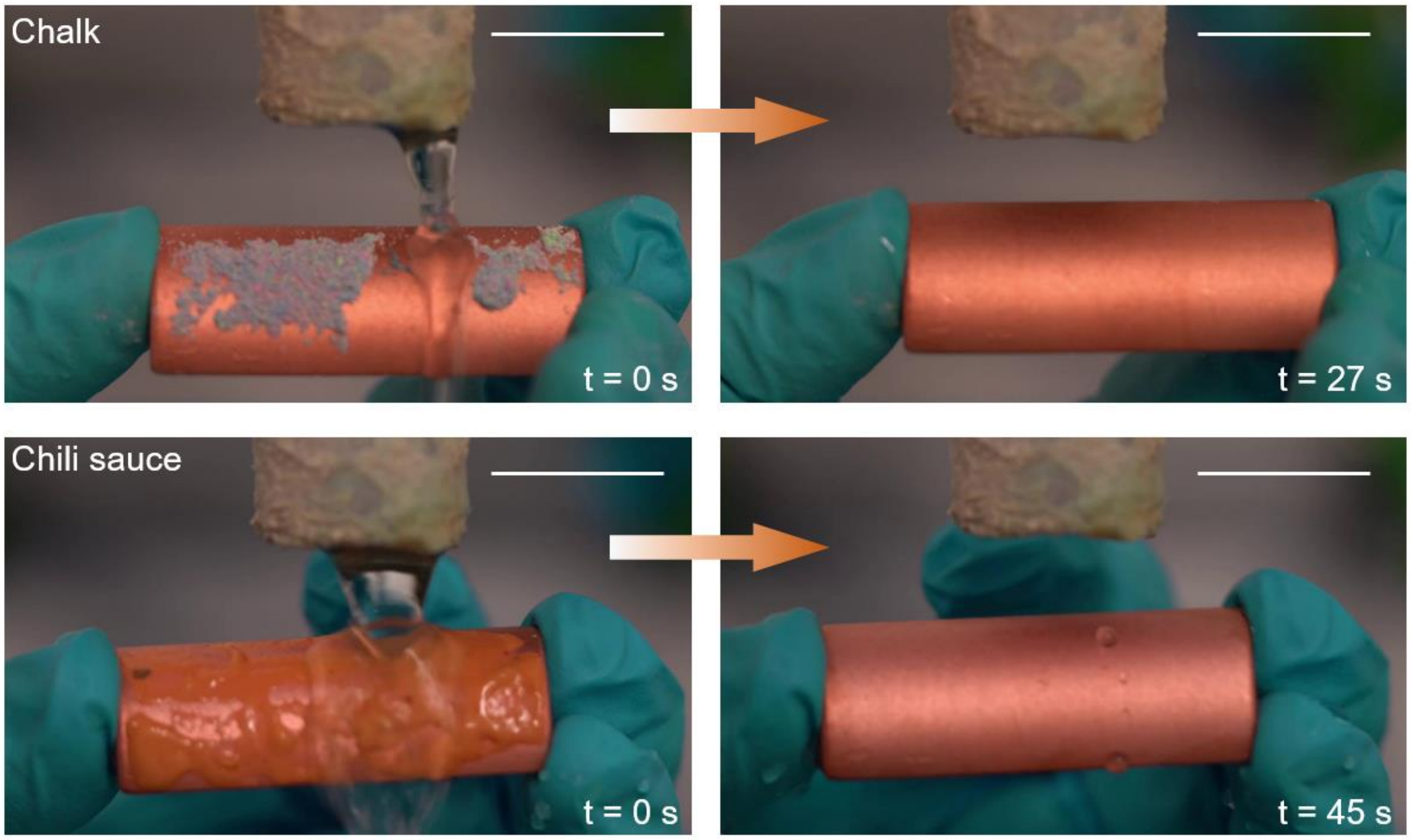

**Figure S10.** Self-cleaning effect. Photographs showing the self-cleaning property of PDMS brushes after being contaminated by chalk powder and chili sauce, when rinsed with tap water. Scale bar: 20 mm.

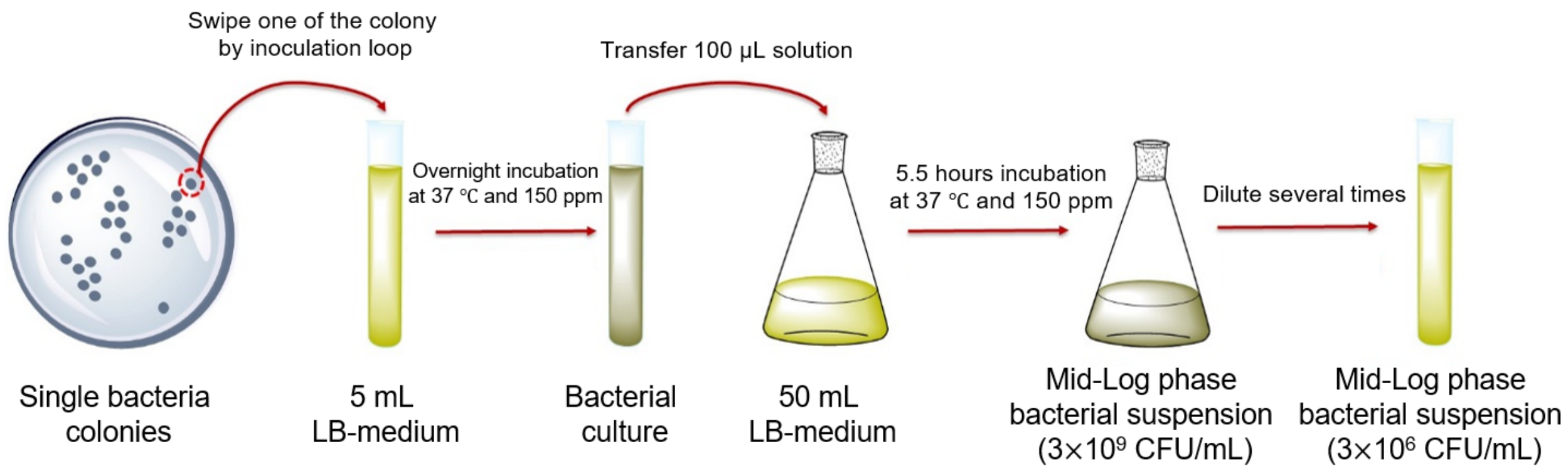

**Figure S11.** Bacterial suspension preparation. Schematic shows the experimental procedure of bacteria culture and solution preparation.

Bacteria adhesion of gram-negative (*Escherichia coli* (*E. coli*), MG1655) and gram positive (*Staphylococcus aureus* (*S. aureus*), DSM 11823) bacteria on the coating surfaces are examined by scanning electron microscopy (SEM, LEO 1530 Gemini, Zeiss). Mid-Log phase bacterial suspension for the test is prepared by the following procedure: a single bacteria colony inoculate into a test tube containing 5 mL LB (Luria-Bertani) medium (2.5% LB (Invitrogen) in Milli-Q water). Then the tube is shaken in a half-covered position in an orbital shaker incubator (Eppendorf™ Innova™ 44) at 37°C and 150 rpm for overnight. Then, 100 µL the overnight bacteria suspension is transferred into 50 mL fresh LB-medium and incubated under stirring, using an incubator at 37°C and 200 rpm for 5.5 hours to reach the mid-Log phase which contained $3x10^9$ colony forming units for per milliliter (CFU/mL). 1 mL mid-Log phase inoculum is diluted several times ($10^1$, $10^2$ and $10^3$) with fresh LB-medium to gain $3x10^6$ CFU/mL to prepare bacteria test suspension.

## Tables

**Table S1.** Thickness of the current state-of-the-art superhydrophobic surfaces.

| *Author/Journal* | *Composition* | *Thickness (μm)* |
|---|---|---|
| Miljkovic, Nenad, et al. *Nano Letters* 13.1 (2013): 179-187. | CuO + tridecafluoro-(1,1,2,2-tetrahydrooctyl)-1-trichlorosilane | 1 |
| Tsoi, Shufen, et al. *Langmuir* 20.24 (2004): 10771-10774. | $SiO_2$ + 3,3,3-trifluoropropylsiloxane | 2 |
| Tang, Yu, et al. *Nano Letters* 21.22 (2021): 9824-9833. | Copper + alkaline oxidation | >8 |
| Wen, Rongfu, et al. *Joule* 2.2 (2018): 269-279. | Copper Nanowire + trichloro (1H,1H,2H,2H-perfluorooctyl)-silane | >15 |
| Wang, Dehui, et al. *Nature* 582.7810 (2020): 55-59. | Armor microstructures + silica nanomaterial | >25 |
| Wu, Shuwang, et al. *Proceedings of the National Academy of Sciences* 117.21 (2020): 11240-11246. | soot particles + tetraethoxysilane + polydimethylsiloxane | 1.5-30 |
| Zhang, Jilin, et al. *Macromolecular Rapid Communications* 25.11 (2004): 1105-1108. | PTFE | 50 |
| Varanasi, Kripa K., et al. *Applied Physics Letters* 97.23 (2010): 234102. | silicon posts + tridecafluoro-1,1,2,2-tetrahydrooctyl trichlorosilane | >100 |
| Peng, Chaoyi, et al. *Nature Materials* 17.4 (2018): 355-360. | fluorinated epoxy + perfluoropolyether | >150 |
| Liu, Yahua, et al. *Nature Physics* 10.7 (2014): 515-519. | Copper + tapered posts + trichloro(1H,1H,2H,2H-perfluorooctyl)silane | 400 |

**Table S2.** Thickness of the current state-of-the-art lubricant infused surfaces.

| *Author/Journal* | *Composition* | *Thickness (μm)* |
|---|---|---|
| Tripathy, Abinash, et al. *ACS Nano* 15.9 (2021): 14305-14315. | Vertical Graphene + Krytox 1525 | 0.07 |
| Preston, Daniel J. et al. *Scientific Reports* 8.1 (2018): 540. | CuO + Krytox GPL 101 | 1 |
| Sett, Soumyadip, et al. *Nano Letters* 19.8 (2019): 5287-5296. | CuO + Krytox 1525 | >1 |
| Sun, Jianxing, et al. *Soft Matter* 15.24 (2019): 4808-4817. | $SiO_2$ + Krytox GPL 102 | >1.5 |
| Li, Shuai, et al. *Langmuir* 38.41 (2022): 12610-12616. | PDMS | >2 |
| Tenjimbayashi, Mizuki, et al. *Langmuir* 34.4 (2018): 1386-1393. | Poly(vinylidene fluoride-co-hexafluoropropylene) + Dibutyl phthalate +Perfluoropolyether | 4.24 |
| Anand, Sushant, et al. *ACS Nano* 6.11 (2012): 10122-10129. | Silicon micro-post + Krytox | 10 |
| Wooh, Sanghyuk, et al. *Angewandte Chemie* 129.18 (2017): 5047-5051. | $TiO_2$ + PDMS | >10 |
| Ge, Qiaoyu, et al. *ACS Applied Materials & Interfaces* 12.19 (2020): 22246-22255. | Etched Aluminum + Krytox 1506 | 20-50 |
| Wong, Tak-Sing, et al. *Nature* 477.7365 (2011): 443-447. | Teflon + perfluoropolyether | 60-80 |

**Table S3.** Cost estimation of PDMS brushes on copper plate. The total cost for PDMS brushes coating is 9.43 USD per m$^2$. The price for the chemicals and electricity can be found from reference.[7-9]

| *Item* | *Price ($)* | *Amount (·1m$^{-2}$ copper)* | *Cost ($)* |
|---|---|---|---|
| Poly(dimethylsiloxane) | 2.22 /kg | 0.97 g | 2.15 |
| Electricity | 0.14 /kWh | | |
| Plasma 120W | | 5min, 0.2 kWh | 0.28 |
| Oven 1250W (max) | | 24h, 1.5 kWh | 0.21 |
| Acetone | 0.86 /kg | 7.89kg | 6.79 |
| Total | | | 9.43 |

**Supporting Videos**

**Supporting Video S**

**Supporting Video S1:** Water drop slide off a PDMS-coated copper tube. Tilt angle: 25°.

**Supporting Video S2:** Water drop sliding on PDMS brushes coated copper plate: $1^{st}$ drop.

**Supporting Video S3:** Water drop sliding on PDMS brushes coated copper plate: $100^{th}$ drop.

**Supporting Video S4:** Water drop sliding on PDMS brushes coated copper plate: $1000^{th}$ drop.

**Supporting Video S5:** Water drop sliding on PDMS brushes coated copper plate: $5000^{th}$ drop.

**Supporting Video S6:** Left: Steam condensation on a hydrophilic copper oxide surface. Right: real-time heat transfer coefficient and subcooling. Steam pressure: 30 mbar. Playback: 240×.

**Supporting Video S7:** Left: Steam condensation on a superhydrophobic aluminium surface. Right: real-time heat transfer coefficient and subcooling. Steam pressure: 30 mbar. Playback: 240×. The video is adapted from our previous work.[10]

**Supporting Video S8:** Left: Steam condensation on PDMS brushes coated copper surface. Right: real-time heat transfer coefficient and subcooling. Steam pressure: 30 mbar. Playback: 240×.

**Supporting Video S9:** Steam condensation on PDMS brushes coated copper surface. Steam pressure: 1.4 bar. Playback: 0.1×.

**Supporting Video S10:** Cleaning process of PDMS brushes coated copper plate contaminated with chalk powder.

**Supporting Video S11:** Cleaning process of PDMS brushes coated copper plate being contaminated with chili sauce.

**Present Addresses:**

#C.W.E.L.: Massachusetts Institute of Technology, Cambridge, MA, 02139, United States

**Supporting References**